\documentclass{aa}

\usepackage{graphicx}
\usepackage{txfonts}
\usepackage{natbib}
\bibpunct{(}{)}{;}{a}{}{,}
\newcommand{\kms}{km\,s$^{-1}$}

\begin{document}

\titlerunning{EK Draconis with PEPSI}
\authorrunning{J\"arvinen et al.}

\title{Mapping EK Draconis with PEPSI\thanks{Based on data acquired with
    PEPSI using the Large Binocular Telescope (LBT).}}

   \subtitle{Possible evidence for starspot penumbrae}

   \author{
S.~P.~J\"arvinen\inst{1}
\and K.~G.~Strassmeier\inst{1}
\and T.~A.~Carroll\inst{1}
\and I.~Ilyin\inst{1}
\and M.~Weber\inst{1}
   }

   \institute{Leibniz-Institut f\"ur Astrophysik Potsdam (AIP),
     An der Sternwarte 16, 14482 Potsdam, Germany\\
              \email{sjarvinen@aip.de}
}

   \date{Received xxx x, 2018; accepted xxx x, 2018}


  \abstract
   {}
   {We present the first temperature surface map of EK\,Dra from
     very-high-resolution spectra obtained with the Potsdam Echelle
     Polarimetric and Spectroscopic Instrument (PEPSI) at the Large Binocular
     Telescope.
}
   {Changes in spectral line profiles are inverted to a stellar surface 
     temperature map using our $i$Map code. The long-term photometric record 
     is employed to compare our map with previously published maps.
}
   {Four cool spots were reconstructed, but no polar spot was seen. The
     temperature difference to the photosphere of the spots is between
     990 and 280\,K. Two spots are reconstructed with a typical solar
     morphology with an umbra and a penumbra. For the one isolated and
     relatively round spot (spot A), we determine an umbral temperature of
     990\,K and a penumbral temperature of 180\,K below photospheric
     temperature. The umbra to photosphere intensity ratio of EK\,Dra is
     approximately only half of that of a comparison sunspot. A test 
     inversion from degraded line profiles showed that the higher spectral 
     resolution of PEPSI reconstructs the surface with a temperature
     difference that is on average 10\,\% higher than before and with smaller
     surface areas by $\sim$10-20\,\%. PEPSI is therefore better suited to
     detecting and characterising temperature inhomogeneities. With ten more
     years of photometry, we also refine the spot cycle period of EK\,Dra
     to 8.9$\pm$0.2 years with a continuing long-term fading trend.
}
   {The temperature morphology of spot~A  so far appears to show the best
     evidence for the existence of a solar-like penumbra for a starspot.
     We emphasise that it is more the non-capture of the true umbral
     contrast rather than the detection of the weak penumbra that is the
     limiting factor. The relatively small line broadening of EK\,Dra, together
     with the only moderately high spectral resolutions previously available,
     appear to be the main contributors to the lower-than-expected spot
     contrasts when comparing to the Sun.
   }

   \maketitle
%

\section{Introduction}

\object{EK\,Draconis} (HD\,129333, G1.5V) is known as probably the best-studied
analogue of the young ($\approx$50~Myr) Sun. Its undepleted lithium abundance,
fast rotation, and strong chromospheric and coronal activity are typical for
the youth of this star. The target also serves as a corner stone for
rotational evolution studies because it is effectively single, bright in
X-rays, and accessible to Doppler- and Zeeman-Doppler-imaging studies. No
planet has been detected around it so far, but that may be just a question of
time. A very good summary of previous observations of this target was given
recently by
\cite{2017MNRAS.465.2076W}
and we refer the reader to this paper.

Solar analogy taught us that the surface spot distribution is a fingerprint
of the underlying dynamo process and its subsequent magnetic-field eruption in
form of sunspots or sunspot groups. On stars, we resolve the surface by an
indirect tomographic imaging technique called Doppler imaging, and map the
surface temperature or brightness distribution as a proxy of the magnetic
field. This technique was introduced to cool stars close to 40 years ago
\citep[][see also the review by \citet{2009A&ARv..17..251S}]{1983PASP...95..565V}.
It requires high-resolution spectra well sampled over a rotation period of the
star but also a target star with rapid rotation so that the line broadening is
dominated by Doppler broadening.

Despite the relatively fast rotation of EK\,Dra with a period of
$\sim$2.6~d, the projected rotational velocity on the stellar equator is
just $\sim$16~\kms . Such small line broadening limits the surface
resolution via the Doppler effect and usually brings the instrumental-profile
width dangerously close to other thermal and velocity broadening mechanisms
like microturbulence
\citep[e.g.][]{1992LNP...397...33C, 1993PASP..105.1415P,2000A&AS..147..151R}.
Simulations have shown that a practical limit for the application of the
Doppler-imaging technique is reached when there are less than five resolution
elements across the projected stellar disk
\citep{1990A&A...233..497P}
or when the exposure time is so long that the rotational drift is of the same
order as the size of the surface feature itself
\citep{1992LNP...397...33C}.
Even with perfect phase coverage and zero microturbulence one can then not
reliably reconstruct features, which makes the term imaging eventually
obsolete. Therefore, the capability of this technique to resolve stellar surface
structure depends on spectral resolution
\citep[see also][]{1993A&A...274..851K, 2003A&A...406..273B}.

The first Doppler image of EK\,Dra was presented by
\citet{DI1}
based on CFHT/Gecko observations with $R=\lambda/\Delta\lambda =120,000$. It
was followed-up with images by
\citet{2007A&A...472..887J, 2009AIPC.1094..660J}
based on NOT/SoFin data with $R\approx77,000$. Instead of temperature maps,
two teams have published brightness maps of EK\,Dra. The brightness maps
presented by
\citet{2016A&A...593A..35R}
were based on TBL/NARVAL observations with $R\approx65,000$. They reveal
larger spots than what is seen in the temperature maps but the locations of
the spots in general, ranging from equatorial regions up to higher latitudes,
are consistent and comparable. The other set of brightness maps was published
just recently by
\citet{2017MNRAS.465.2076W}
based on combinations of CFHT/ESPaDOnS ($R\approx68,000$) and TBL/NARVAL
observations. The brightness maps from 2007, produced by the two teams, that is 
\citet{2016A&A...593A..35R} 
and 
\citet{2017MNRAS.465.2076W},
are similar at the low latitudes, but the map using the
first part of the 2007 observations by
\citet{2017MNRAS.465.2076W}
shows a prominent high-latitude feature. The latter may have been present also
in the map by
\citet{2016A&A...593A..35R},
but if so, at somewhat lower latitude. Similarly, the brightness maps from
observations in 2012 agree well with each other at low latitudes, but the
polar region in the map by
\citet{2016A&A...593A..35R}
is featureless while the map by
\citet{2017MNRAS.465.2076W}
shows a small but significant polar spot. Such discrepancies are not
surprising because at the resolving power of $R =70,000$ (4.3~\kms) one has
merely eight resolution elements across the stellar disk. Combined with a
non-perfect phase coverage, artefacts will arise from this limitation.

In this paper, we present the first Doppler image employing the
very-high-resolution capability of the Potsdam Echelle Polarimetric and
Spectroscopic Instrument (PEPSI) at the 2$\times$8.4\,m (effective aperture of
11.8m) Large Binocular Telescope (LBT). This enabled a spectral resolution of
up to $R\approx 250,000$, which translates to approximately 25--30 resolution
elements across the stellar disk of EK\,Dra. The observations are briefly
described in Sect.~2. Our Doppler imagery is presented in Sect.~3, while
Sect.~4 discusses the results and presents our conclusions.

\section{Observations and data reduction }\label{sect:obs}

\subsection{High-resolution spectroscopy}

\begin{table}
  \caption{Logbook of the PEPSI observations. }
\label{T:obs}
\centering
\begin{tabular}{c c c c c r }
\noalign{\smallskip}\hline\hline\noalign{\smallskip}
Date  & HJD          & Phase & S/N & S/N & S/N \\
(UTC) & 2,450,000+ &       & CD\,III & CD\,V & aver\\
\noalign{\smallskip}\hline \noalign{\smallskip}
2015/04/03 & 7115.8866 & 0.205 & 181 & 287 & 5658 \\
2015/04/09 & 7121.7174 & 0.442 & 111 & 189 & 3573 \\
2015/04/09 & 7121.9181 & 0.519 & 197 & 313 & 6420 \\
2015/04/09 & 7122.0120 & 0.555 & 238 & 382 & 6559 \\
2015/04/10 & 7122.7021 & 0.820 & 207 & 244 & 5421 \\
2015/04/10 & 7122.8576 & 0.880 & 250 & 394 & 8781 \\
2015/04/10 & 7122.9744 & 0.925 & 239 & 383 & 7341 \\
2015/04/11 & 7123.6768 & 0.193 & 165 & 196 & 5000 \\
2015/04/11 & 7123.8599 & 0.264 & 275 & 433 & 10627 \\
2015/04/11 & 7123.9413 & 0.296 & 226 & 361 & 6444 \\
\noalign{\smallskip}\hline
\end{tabular}
\tablefoot{The first column gives the date, the second column the heliocentric
  Julian date (HJD), the third column the rotational phase based on the
  ephemeris given in Eq.~\ref{eq1}, and the last three columns give the 
  average S/N per pixel for the two wavelength regions of CD~III and V
  and the S/N of the weighted average line profile.}.
\end{table}

The spectroscopic observations in this paper were obtained with PEPSI at the
2$\times$8.4\,m LBT in Arizona. We employed PEPSI's $R =250,000$ mode with
seven-slice image slicers and 100$\mu$m fibres. Spectral resolution varies
with wavelength (higher in the red, lower in the blue), with position in the
\'echelle order (higher in the centre), and camera-focus position across the
CCD 
\citep[for more details, see][Fig.~3 in both papers]{2017Strassmeier,2017PEPSI}.
Spectrograph focus during these observations was such that the full width
at half maximum (FWHM) of the Th-Ar comparison lines had an average spectral
resolution of $230,000 \pm 30,000$. The instrument is described in detail by
\citet{2015AN....336..324S}.

Monitoring of EK\,Dra continued over eight nights as part of the instrument
commissioning in April 2015. Ten spectra were taken in two wavelength
bands simultaneously with cross disperser (CD)~III covering 4800--5441~\AA\ 
and CD~V covering 6278--7419~\AA. Exposure time was 10~min and typically two 
back-to-back exposures were obtained and average combined. Average 
signal-to-noise ratio (S/N) was around 250 per pixel in CD~III and 350 in 
CD~V. The logbook of the observations is given in Table~\ref{T:obs}. Besides 
the phase-resolved spectra, we also obtained one spectrum in the remaining 
four CDs of PEPSI. Together with the two CDs from the monitoring, we added 
these to one phase-averaged (deep) spectrum of EK\,Dra covering the entire 
PEPSI format (3837--9140\,\AA).

Data reduction was done with the software package SDS4PEPSI (``Spectroscopic
Data Systems for PEPSI'') based on
\citet{4A},
and described in more detail in
\citet{2017Strassmeier}.
It relies on adaptive selection of parameters by using statistical inference
and robust estimators. The standard reduction steps include bias overscan
detection and subtraction, scattered light extraction from the inter-order
space and subtraction, definition of \'echelle orders, optimal extraction of
spectral orders, wavelength calibration, and a self-consistent continuum fit
to the full two-dimensional (2D) image of extracted orders.

\subsection{Photometry}

The new photometry was taken with the T7 Amadeus telescope, one of the two
0.75m Vienna-AIP automatic photoelectric telescopes (APT) at Fairborn
Observatory in southern Arizona. Measurements were made differentially between
the variable star, a comparison star (HD\,129390), a check star (HD\,129798),
and a sky position using three ten-second integrations. The telescope 
and photometer themselves are described in
\citet{1997PASP..109..697S},
while the automatic data reduction is described in
\citet{1997A&AS..125...11S}
and
\citet{2001AN....322..325G}.
A total of 1138 new Johnson $V$-band data points are presented covering HJD 
range 2,454,517--2,457,910 (February 2008 -- June 2017).

\section{Doppler imaging using PEPSI data}\label{sect:di}

\subsection{Assumptions}

\begin{table}
\caption{Relevant astrophysical properties of EK\,Dra. } \label{T:ekdra}
\begin{small}
\begin{tabular}{lll}
\hline \hline \noalign{\smallskip}
Parameter                   & Value   & Based on  \\
\noalign{\smallskip}\hline \noalign{\smallskip}
Classification, MK          & G1.5V  & S\&R (1998) \\
Effective temperature, K    & 5,730$\pm$50 &  spectrum synthesis  \\
                            & 5,750        &  Doppler imaging \\
Log gravity, cgs            & 4.41$\pm$0.03  & spectrum synthesis \\
$v\sin i$, \kms             & 17.5$\pm$0.2 & spectrum synthesis \\
                            & 16.6$\pm$0.2 & Doppler imaging \\
Microturbulence, \kms       & 2.0$\pm$0.1 & spectrum synthesis \\
Macroturbulence, \kms       & 4.0 & assumed \\
Rotation period, d          & 2.606$\pm$0.001 & APT $V$ photometry \\
Inclination, deg            & 63$\pm$2 & Doppler imaging \\
Metallicity, [Fe/H]$_\odot$  & --0.2$\pm$0.02  & spectrum synthesis   \\
Chemical abundances         & solar & spectrum synthesis \\
\noalign{\smallskip}\hline
\end{tabular}
\end{small}
\tablebib{S\&R (1998) = \citet{DI1}}
\tablefoot{Values not cited in the third column were obtained in this paper.}
\end{table}

The main assumption comes from knowing the immaculate spectral line profile of
the star, that is, the line profile without the surface inhomogeneities that 
one wants to reconstruct. Our procedure is based on prior local thermodynamic
equilibrium (LTE) spectrum synthesis in a wavelength range that is as large as
possible with exclusion of spectral regions that are known to be prone to
magnetic activity. Besides resonance lines, this excludes optically thick
lines, that is, strong and saturated lines with possible chromospheric and/or
non-LTE contaminations. In cool stars, also the band heads of TiO and VO
molecular series are excluded because of likely (cool) spot contamination. The
resulting fit values for effective temperature, gravity, metallicity, and
turbulence are then the starting values for our line-profile inversion.

The stellar parameters of EK\,Dra were first verified with the spectrum
synthesis code ParSES
\citep[e.g.][]{2006ApJ...636..804A}
and model atmospheres from MARCS
\citep{2007ASPC..378...60G}.
We adopted the \emph{Gaia}-ESO clean line list
\citep{2014A&A...564A.133J}
with various mask widths around the line cores between $\pm$0.05 and
$\pm$0.25\,\AA . As target spectrum, we used the average-combined spectrum
from the entire PEPSI data set. This deep spectrum covers the full wavelength
range of 3837--9140\,\AA. Its S/N is inhomogeneous though because the
Doppler-imaging monitoring covered ten spectra in CD~III+V while the other CDs
had only a single observation. The S/N in the CD~V, III, VI, and IV ranges
therefore peaks at 1000, 600, close to 400, and 250, respectively, and in the
bluest part between 100 and 200 (which, however, are not used here). The best
ParSES fit leads to $T_{\rm eff}$=5,730$\pm$50\,K, $\log g$=4.41$\pm$0.01,
[M/H]=$-$0.2$\pm$0.02, $\xi_t$=2.0$\pm$0.1~\kms, and
$v \sin i$=17.5$\pm$0.2~\kms. A recent application to an equivalent PEPSI
spectrum of the Sun-as-a-star revealed good agreement with canonical solar
values except maybe for $v\sin i$ where ParSES preferentially converged on
$v\sin i=0\pm1$\,\kms\ and a microturbulence of 1.2$\pm$0.2\,\kms\
\citep[see also][]{2017PEPSI}.
From comparative studies, we would have expected a solar $v\sin i$ of 1.9 and a
microturbulence of 1.0 or slightly less. It appears that there is remaining
cross talk between these two broadening mechanisms at the sub-\kms\ level.
However, this will affect the values for EK\,Dra only at a negligible level
because its $v\sin i$ $\approx$17\,\kms\ is far beyond the solar value. 

Another critical input parameter is the stellar rotation period itself. This is
usually best determined from contemporaneous phase-resolved photometry.
However, EK\,Dra's photometric period appears to change with time. Values
between 2.55 and 2.89\,d were measured in the past and explained due to spots
appearing at different latitudes at different times on a differentially
rotating surface (see later Sect.~\ref{S4}). In the present paper, we
incorporate differential rotation implicitly in the inversion process by
allowing for a latitude-dependent rotation period (actually relative angular
velocity). Our assumption is that the angular velocities vary with latitude,
$\theta$, only as $\sin^2\theta$, like on the Sun. We adopt the observed,
disk-averaged photometric period of 2.606\,d as a starting value.

\citet{1994ApJ...428..805D}
estimated the inclination of the rotational axis to be 60\degr\ from the
photometric period, an average $v\sin i$, and an assumed stellar radius of
0.92\,R$_\sun$. Line-profile inversions by
\citet{DI1}
with inclinations between 20\degr\ and 90\degr\ did indeed result in better
fits for $i > 55$\degr\ but did not reveal a particular significant minimum
above that limit. This is basically because the Doppler-imaging technique
becomes increasingly insensitive to inclination effects once $i$ is above
60\degr. This exercise was repeated with the current data leading to a refined
value of 63\degr$\pm$2\degr.

The well-known bisector shape from surface granulation
\citep[e.g.][]{2008A&A...492..199D},
which $i$Map treats in a global manner as a microturbulence broadening in
plane-parallel model atmospheres, may in principle also add to the overall
residuals. However, the bisector averaged from many hundred lines in the deep
spectrum did not show a conclusive deviation from a straight line.

Table~\ref{T:ekdra} summarises the adopted input parameters for our Doppler
images.

At this point we note that the Hubble Space Telescope (HST) has witnessed
EK\,Dra having one of the largest far-ultra-violet (FUV) flares ever recorded
on a sun-like star
\citep{2015AAS...22513823A}.
Generally, the HST observations show that the outer atmosphere of this star
is very complex, energetic, and dynamic
\citep{2010ApJ...723L..38A, 2015AAS...22513823A}.
Care must be exercised not to include flare-affected optical spectra for the
mapping
\citep[see][]{2017A&A...597A.101F, 2017Strassmeier}.
No flares were seen in our contemporaneous optical photometry.

\subsection{Inversions}

   \begin{figure*}
   {\bf a.}\\
   \includegraphics[width=\textwidth]{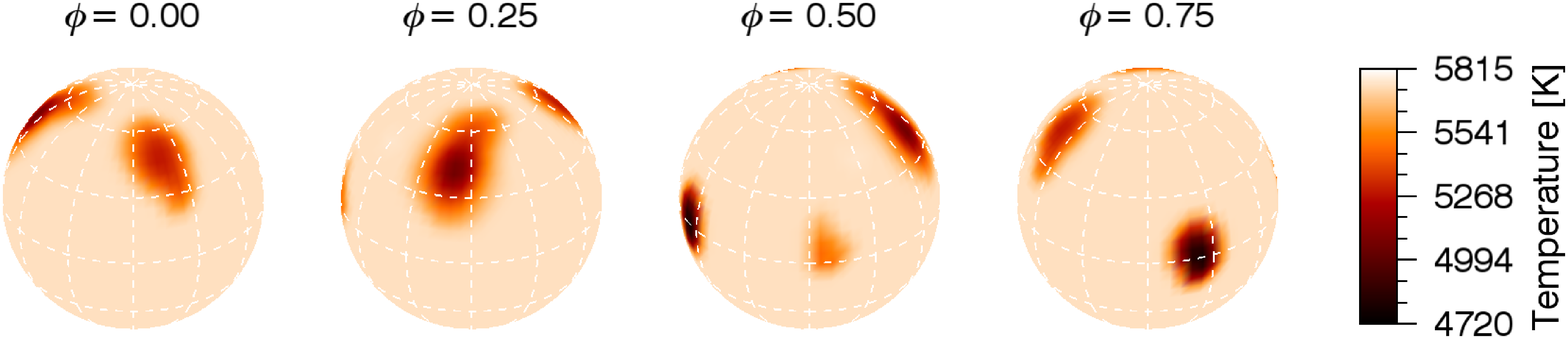}
   {\bf b.}\\
   \includegraphics[width=\textwidth]{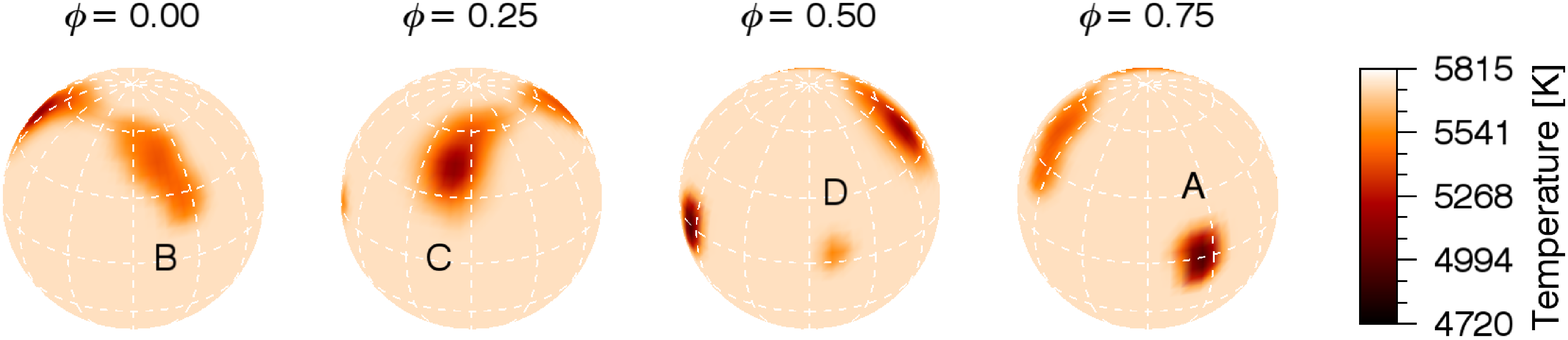}
   \caption{Temperature Doppler images from PEPSI observations.
     \emph{Panel a.} Based on the original data with a spectral resolution
     of $R$=230,000 on average.
     \emph{Panel b.} Based on the same data but downgraded to a spectral
     resolution of 65,000. The spots are labelled A to D.}
         \label{di}
   \end{figure*}

   \begin{figure}
   \centering
   \includegraphics[width=.5\textwidth]{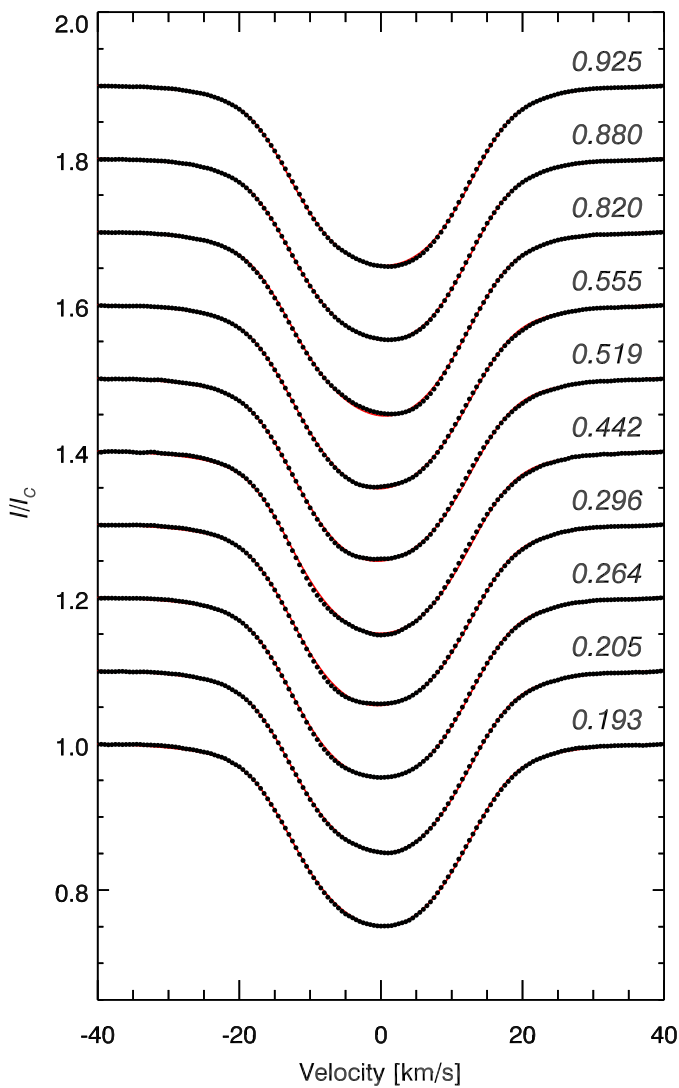}
   \caption{Observed (black dots) and computed (red lines) line profiles
     for the high-resolution Doppler image in Fig.~\ref{di}a. Profiles are
     labelled with their respective rotational phases. Rotation advances from
     bottom to top.
              }
         \label{lineprof}
   \end{figure}

   \begin{figure}
   \centering
   \includegraphics[width=.5\textwidth]{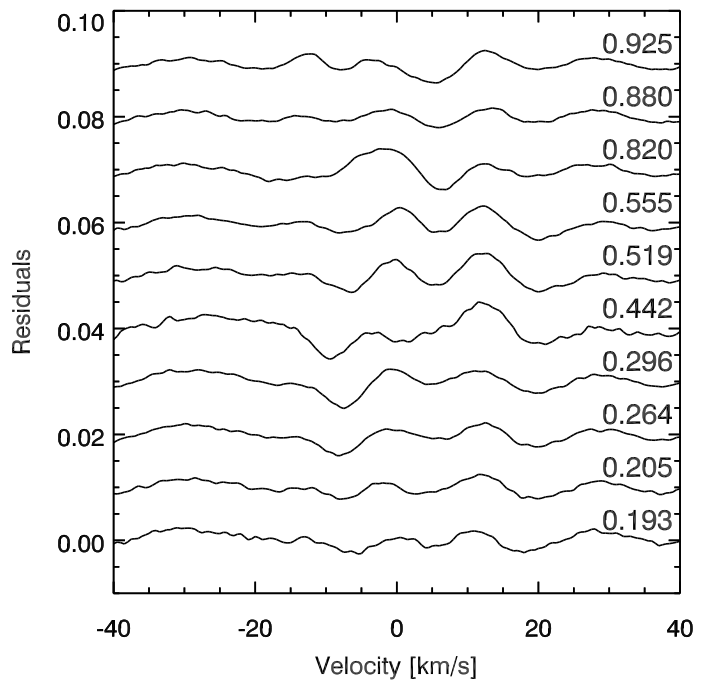}
   \caption{The line profile residuals (observed - inverted). As in
     Fig.~\ref{lineprof}, the residual profiles are labeled with their
     respective phases. The phase-combined rms residual is $4.2\times10^{-4}$
     but systematic deviations appear in individual profiles at
     $\pm 3\times10^{-3}$. We note that the S/N of the line-averaged data is
     between 3,500 and 10,000 and is thus in the same range as the residuals.
              }
         \label{profdiff}
   \end{figure}

   \begin{figure}
   \centering
   \includegraphics[width=.47\textwidth]{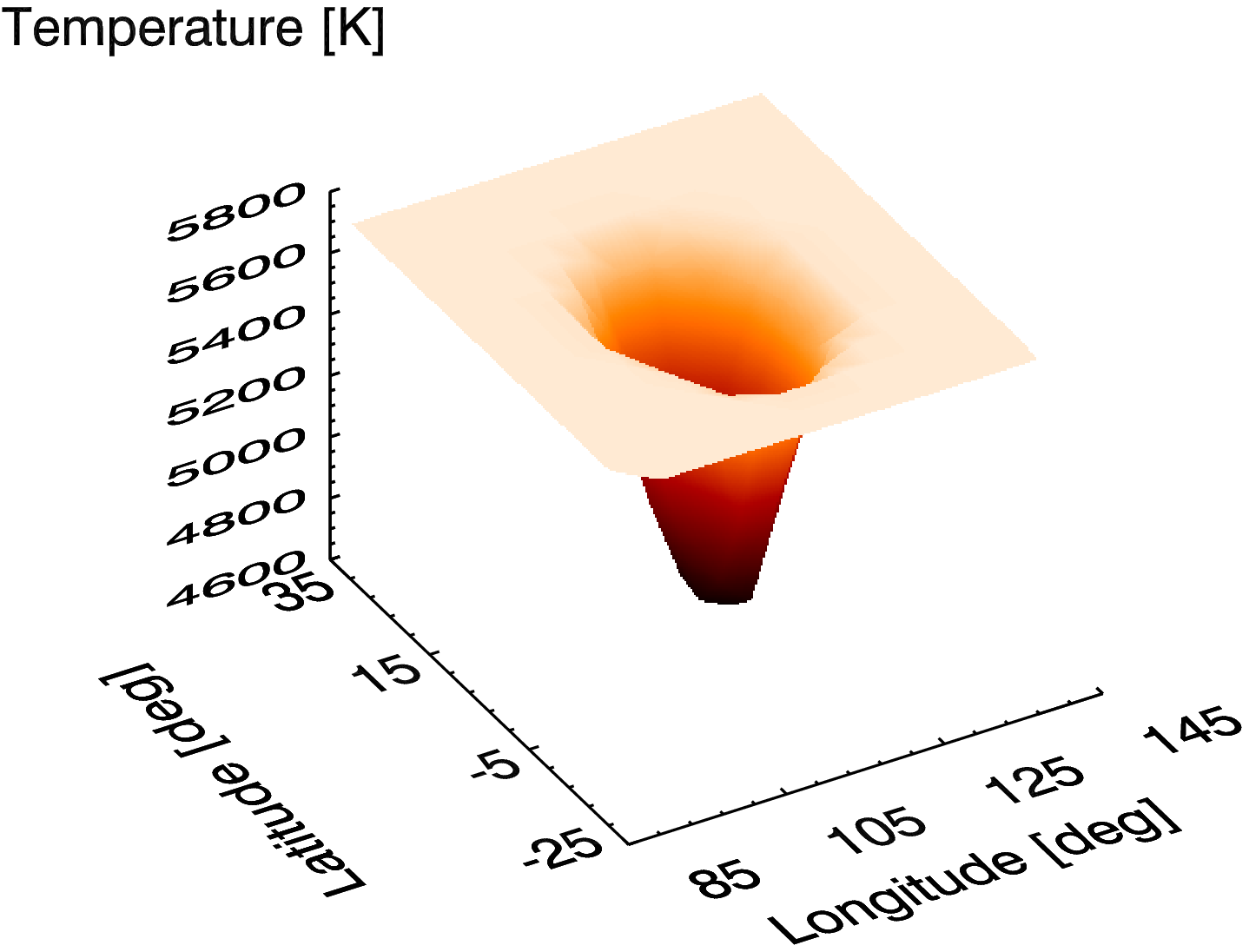}
   \includegraphics[width=.47\textwidth]{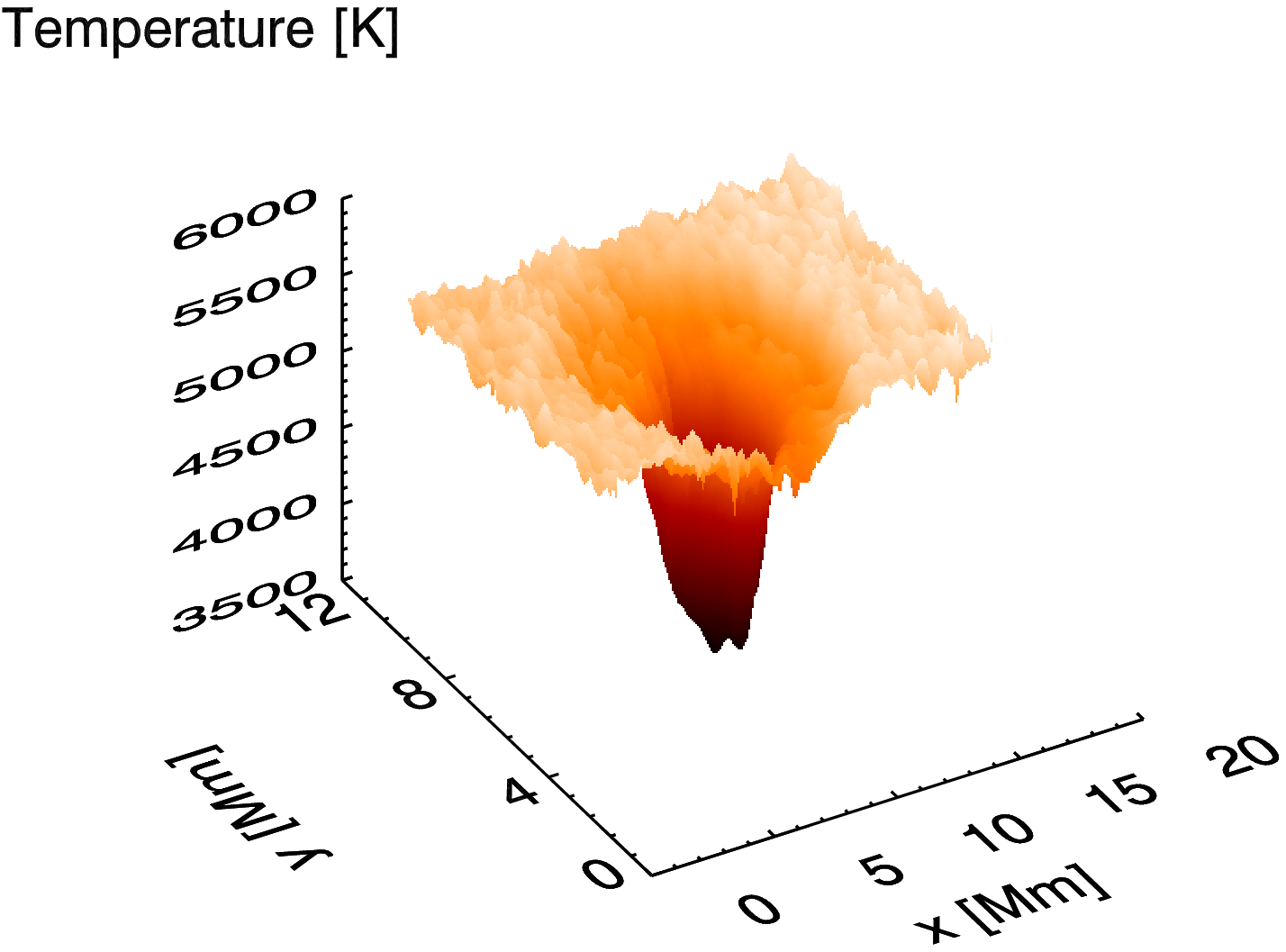}
   \includegraphics[width=.47\textwidth]{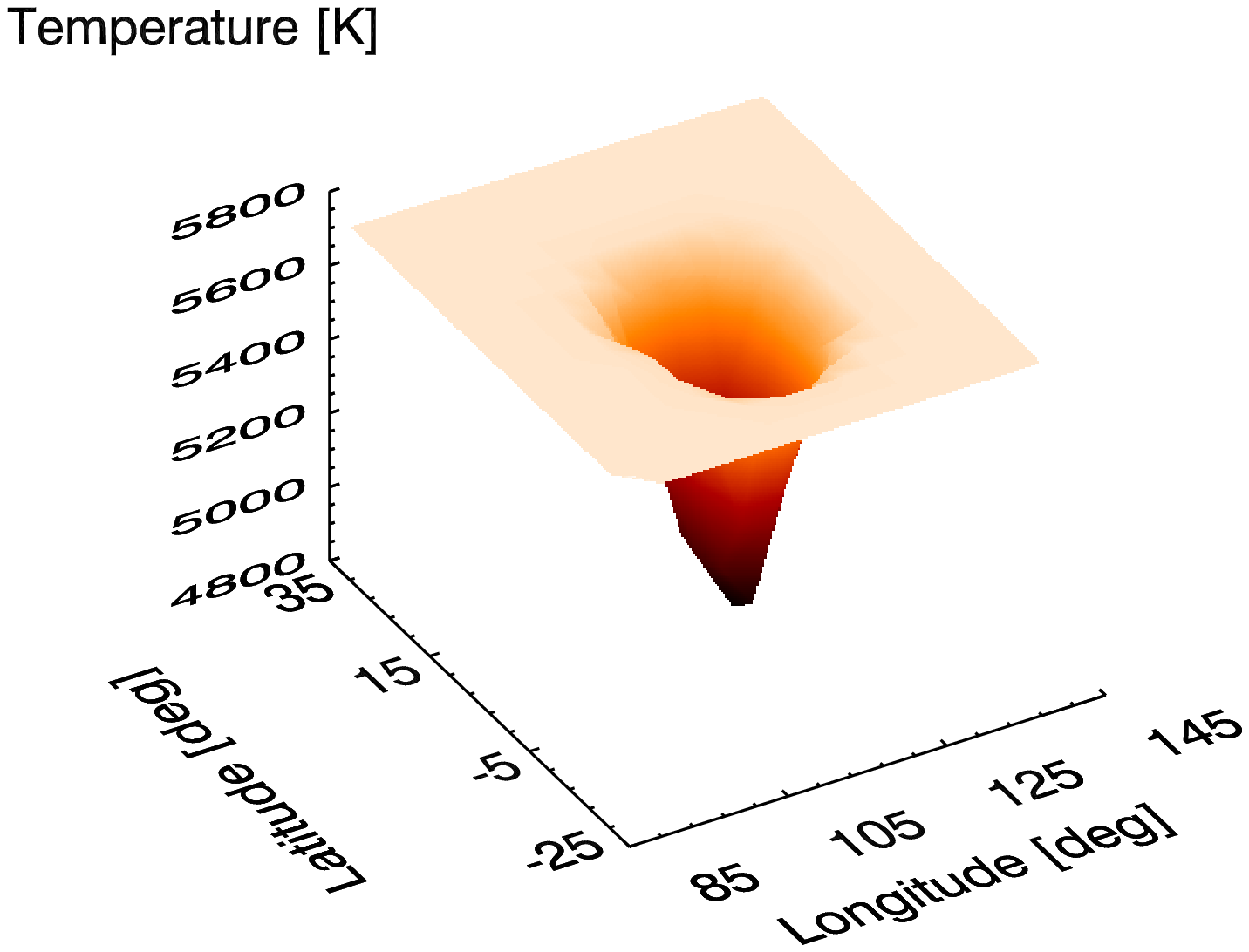}
   \caption{Morphology of spot A (top) compared to a sunspot
     (middle). The bottom plot shows the morphology of spot A when
     low-resolution data are used. The difference is the weaker umbral region.
     The x-axis has spot latitude, y-axis has longitude, and z-axis has the
     temperature elevation. The x- and y-axes of the sunspot
     \citep[adapted from][]{sunspot}
     are in Megameters.
              }
         \label{Fig:spot}
   \end{figure}

The stellar surface is reconstructed using the $i$Map code
\citep[for details, see][]{2007AN....328.1043C, 2009IAUS..259..633C, 2012A&A...548A..95C}.
The code can either perform multi-line inversions for a large number of 
photospheric line profiles simultaneously or use a single average line 
profile. For the present application, we used a weighted average of 207 
spectral lines from wavelength ranges 5000--5400~\AA\ and 6300--7400~\AA\ 
with line depths larger than 60\% (CD~III) and 40\% (CD~V) of the continuum.
Weaker lines were excluded because they decrease rather than increase the 
S/N. However, for CD~V the limit had to be relaxed because there were not
enough lines left after applying the original cut-off. The weighted average
profiles have S/Ns of $\sim$3,500--10,000 per pixel. A total of ten
rotational phases fairly equally distributed are available for the inversion.

The rotational phases were calculated from the ephemeris in Eq.~(\ref{eq1}),
\begin{equation} \label{eq1}
HJD = 2,445,781.859 + 2.606 \times\ E \ ,
\end{equation}
where the period is the average photometric period taken from
\cite{2005A&A...440..735J}
and the zero point is adopted to be a time of the beginning of the photometric
record. We note that a photometric period always reflects the latitude
distribution of spots that appeared contemporaneously on the surface of the
star. If the star is like the Sun and a differentially rotating body, the true
rotation period of the equator will be slightly faster. For high-precision
data with high surface resolution, as in this paper, differential rotation
may no longer be negligible. Therefore, our Doppler imagery will be done in
a comparative way, that is, once with differential surface
rotation and once without.

For the line profile computation, $i$Map solves the radiative transfer with the
help of an artificial neural network. The atomic parameters for the line
synthesis are taken from the Vienna Atomic Line Database
\citep[VALD;][]{2011BaltA..20..503K, 2015PhyS...90e4005R}.
These are used with a grid of Kurucz ATLAS-9 model atmospheres
\citep{2004astro.ph..5087C}
to compute local line profiles in 1D and in LTE. The grid covers temperatures
between 3,500\, and 8,000\,K in steps of 250\,K interpolated to the gravity,
metallicity, and microturbulence from the synthesis fits.

For the numerical integration, the stellar surface is partitioned into
5\degr~$\times$~5\degr\ segments, resulting in 2,592 surface segments for the
entire sphere. At the average resolving power of
$\lambda/\Delta\lambda =230,000$, that is, 1.3~\kms, and an average full width
of the lines at a continuum level of $2 \ (\lambda/c) \ v\sin i$ = 0.7~\AA\
(0.54--0.84\,\AA\ between 4800 and 7400\,\AA), we have, on average, an
unprecedented 27 resolution elements across the stellar disk. The spherical
integration grid samples one spectral resolution element at least with two
surface segments when near the stellar equator and near the central meridian.
This also means that the integration time shall not exceed 20\,min, meaning that
phase smearing remains significantly sub-pixel and thus negligible (as a
comparison, rotational phase smearing during a 50-min integration as in our
CFHT/Gecko data from 1995 amounted to 0\fp 013, i.e. of the order of a
resolution element of PEPSI).

\subsection{Results}

Our surface reconstruction of EK\,Dra in Fig.~\ref{di}a reveals three large
spots or spot groups and one small feature. These four spots are labelled A--D
in Fig.~\ref{di}b, and their parameters are quantified in Table~\ref{T:spots}.
The spot centres and areas are defined using isothermal contours as described,
for example, by 
\citet{XXTri}.
No warm spots are reconstructed and also no strictly polar feature is seen.
The coolest spot, spot~A, with a temperature difference of $\Delta T = 990$~K
(between unspotted photosphere and spot core) is located on the stellar
equator at a phase $\phi$=0.65--0.73. It appears well isolated from the
other spots in longitude and latitude. The two larger spots, spots~B and C,
are at mid-latitudes, ranging approximately from 30\degr\ to 60\degr\ and
appearing with a complex elongated shape. Spot~C at $\phi$=0.20--0.30 has a
central temperature that is $\sim$700~K cooler than the unspotted surface
while spot~B at $\phi$=0.85--1.0 appears without any temperature gradient
and is only 470~K cooler than the surrounding. The fourth and smallest spot,
spot~D, is located on the equator at a phase of $\phi$=0.47 with a temperature
difference of only $\sim$280~K.

The best line profile fits achieved a $\chi^2$ of $4.2\times10^{-4}$ and are
compared with the observations in Fig.~\ref{lineprof}. This fit approaches the
S/N of the weighted average data but does show residuals in the line core 
with a peak amplitude of $\pm 3\times10^{-3}$. We believe these deviations 
(Fig.~\ref{profdiff}) relate to the summed impact from several slightly 
ill-determined stellar parameters. Other subtle effects like the use of 
plane-parallel atmospheres in LTE or the magnetic line broadening may 
contribute at this level.
\citet{2017MNRAS.465.2076W}
has shown that EK\,Dra has a magnetic field with up to approximately
$\pm$200\,G which has not been taken into account in the local line profiles.

The photospheric temperature was assumed to be the effective temperature and
was fixed to 5,750~K, which resulted in the overall best fits. From repeated
inversions with slightly deviating parameter combinations like, for example,
decreased gravity and lowered metallicity (but with always the same line
list), we found that the recovered spot temperatures remained surprisingly
stable at the $\pm$50\,K level. If the parameters become grossly different
from what is expected, the code introduces easily recognisable systematic
artefacts like dark or bright circum-stellar rings or simply cheats to find a
suitable image. We cannot assign an absolute error to each spot's temperature
but state that the errors are likely smaller than 50\,K.

In order to quantify the gain from the very-high spectral resolution, we have 
convolved the PEPSI data with a Gaussian of width proportional to 
$\lambda/\Delta\lambda =65,000$, a value used for previous EK\,Dra mappings.
However, this does not change the sampling of the data, which means it 
is an optimistic comparison in favour of the lower-resolution spectra.
These profiles are then inverted in the same way as the original profiles. The
resulting map is presented in Fig.~\ref{di}b where it can be compared directly
with the original map. Its $\chi^2$ for the best fit is $3.2\times10^{-4}$,
thus very comparable. The location of the spots are reconstructed almost
identically to the high-resolution map; in particular the latitudes, which
are typically the quantities the most prone to artefacts. The main difference,
however, is that the central (umbral) temperatures of the three larger spots
are about 100~K warmer from the lower-resolution spectra compared to the map
from the higher-resolution spectra. This translates to a full 10\%. The
weakest spot, spot~D, is also found warmer from the lower-resolution spectra,
but its temperature is only $\sim$50~K higher instead of $\sim$100~K.
The second notable difference is that the high-latitude spots B and C appear
larger by 19\% and 5\%, respectively. This causes the two spots to appear
almost merged at high latitudes whereas the very-high resolution data led to
clearly separated spots. The effect is the opposite for the two small
low-latitude spots; spot A appears smaller by 13\%, spot~D smaller by 33\%
from the lower-resolution spectra.

\begin{table}
  \caption{Spots on EK\,Dra in April 2015.}
\label{T:spots}
\centering
\begin{tabular}{c c r r r r c}
\hline\hline\noalign{\smallskip}
Spot  & $\phi$ & Lon & Lat & $\Delta T_{\rm umbra}$ & $\Delta T_{\rm penumbra}$ & Total area \\
ID    & & (\degr )  & (\degr ) & (K) & (K) & (\% ) \\
\noalign{\smallskip}\hline\noalign{\smallskip}
A & 0.67 & 115 &  3 & 990 & 180   & 5 \\
B & 0.94 &  20 & 48 & 470 & \dots & 8 \\
C & 0.27 & 260 & 43 & 700 & 230   & 12 \\
D & 0.47 & 190 &  8 & 280 & \dots & 3 \\
\noalign{\smallskip}\hline
\end{tabular}
\tablefoot{Longitudes and latitudes are given for the spot centres. The total
  spot area is given as a percentage of the visible hemisphere.}
\end{table}

\section{Discussion and conclusions}\label{S4}

\subsection{Differential surface rotation}\label{S41}

Significant differential surface rotation was recently discovered from
Stokes~$V$ ZDI observations by
\citet{2017MNRAS.465.2076W}.
Puzzlingly, they had seen no evidence for it from Stokes~$I$. This prompted
us to also search for differential rotation (DR) in the (Stokes~$I$) PEPSI data
set by applying $i$Map in its smeared image version. No DR was found in
agreement with
\citet{2017MNRAS.465.2076W}
although we want to emphasise that almost all of our observations are 
obtained within one stellar rotation of EK\,Dra. The question therefore
remains as to why one sees DR from Stokes~$V$ but not from Stokes~$I$.

We note that the shortest equatorial period, $P$=2.51\,d, given by
\citet{2017MNRAS.465.2076W}
from Stokes~$V$ leads to an $i$Map Stokes~$I$ map where the high-latitude 
spots merge and the smallest spot (D) disappears. Another $i$Map inversion 
of our data phased with the longest period from the Stokes~$V$ DR fits, 
$P$=2.766\,d, leads to a comparable map to our test map in Fig.~\ref{di}b 
(which used 2.606\,d) but with spot (D) hardly detected. Furthermore, the 
high-latitude spots appear more elongated than in our 2.606\,d map while the 
central spot regions were not reconstructed to be as cool as in previous
maps. For comparison of the fit qualities, the $P$=2.606\,d map gave a
$\chi^2$ of $4.2\times10^{-4}$ whereas $P$=2.51\,d gave a
$\chi^2$ of $5.4\times10^{-4}$ and $P$=2.766\,d gave $5.0\times10^{-4}$,
which we consider to be significantly worse than the 2.606-d inversion.

Support for the non-detection of DR from Stokes~$I$ comes from our
contemporaneous photometry. For all data sets, $P$=2.606\,d works reasonable
fine, whereas $P$=2.51\,d and $P$=2.766\,d do not produce recognisable phase
curves at all. We note however that for most observing seasons the entire
seasonal data cannot be phased together because of intrinsic changes. A full
observing season typically lasts several months. On average we see that the
light curves remain stable for one month.

\subsection{Spot morphology}

   \begin{figure}[!ht]
   \centering
   \includegraphics[width=.45\textwidth]{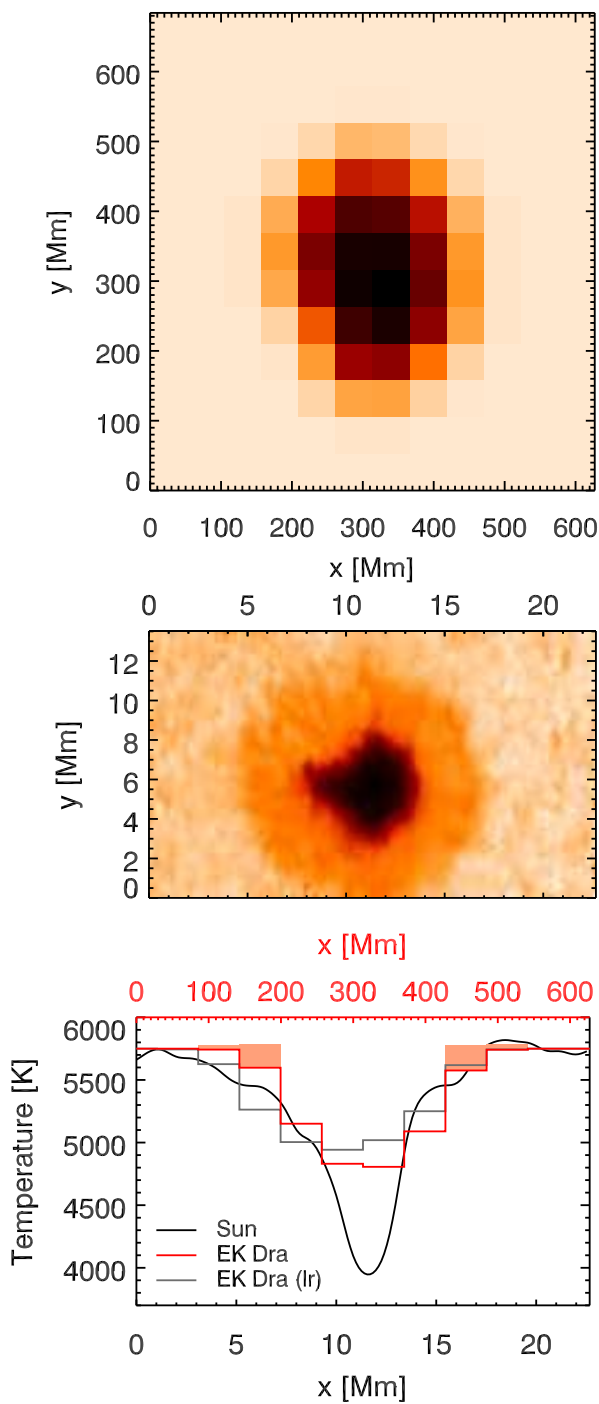}
   \caption{
     \emph{Top panel:} $i$Map reconstruction of spot A. Pixel size is
     5\degr\ both in longitudinal and in latitudinal direction. The scaling
     of 27:1 from surface pixels to Megameters is based on a stellar radius
     of $0.98$\,R$_\sun$.
     \emph{Middle panel:} Image of an isolated sunspot
     \citep[adapted from][]{sunspot}.
     \emph{Bottom panel:} Temperature distribution of the sunspot (black
     line) in comparison to the mean temperature distribution of spot A of
     EK\,Dra (red histogram) and to the low-resolution (lr) result of spot
     A (grey) histogram). The bins belonging to the penumbra of EK\,Dra are 
     indicated with light orange colour (shifted by +100\,K for better 
     visibility) while the extent of the sunspot penumbra can be directly 
     compared with the middle panel. 
}
         \label{f:sunspot}
   \end{figure}

The dominating equatorial spot, spot~A, appears isolated from the other
spotted regions. Therefore, if at all, it is at least minimally affected by
cross talk from other spots during the line profile inversion. We chose it for
a more detailed analysis. Its morphology (radial temperature profile) is
illustrated in Fig.~\ref{Fig:spot}. Although the coolest temperature in the
centre is covered by only one surface pixel, its neighbouring pixels reach
almost the same temperature contrast of $\sim$1,000~K. We therefore accept
this minimum with confidence. This central region in the spot could be termed
its umbra in analogy to sunspot morphology. Consequently, the outer parts
could be dubbed the penumbra, appearing on average with an (effective)
temperature contrast of only 180~K. This is the average of the temperature of
the spot's outer pixel ring compared to the neighbouring photosphere.

While the solar umbra to photosphere intensity ratio is around 0.25, the solar
penumbra to photosphere intensity ratio is around 0.80
\citep[e.g.][]{1964suns.book.....B}.
In terms of effective temperature (at optical depth of 0.67 instead of 1.0 as
for the radiation temperature) typical solar penumbrae are cooler by 270~K and
umbrae cooler by 1,600--2,000~K than the photosphere. Compared to a typical
sunspot the umbra-to-penumbra contrast on EK\,Dra appears to be only half of
this, as demonstrated in Fig.~\ref{f:sunspot}. The lower panel of 
Fig.~\ref{f:sunspot} compares  the radial temperature profiles of spot A 
(reconstructed with two different resolutions) and a Sunspot. The
penumbra of spot A is highlighted by the shaded area. We note that in this 
figure the temperature profiles of both the Sun and EK\,Dra match the same 
effective temperature but the EK\,Dra spot is 27 times larger than the sunspot
with a diameter of $\sim$400\,Mm based on a stellar radius of EK\,Dra of
$0.98$\,R$_\sun$; the latter is based on the \emph{Gaia} parallax of 27.90\,mas
\citep{2016A&A...595A...2G}.
There are no such large spots on the Sun. Nevertheless, the stellar spot
appears much too warm in its centre compared to its solar counterpart.

We note that spot~C in our map in Fig.~\ref{di}a also shows an umbra/penumbra
morphology but with an even shallower temperature gradient than spot~A. Its
central temperature is only 700\,K cooler than the photosphere, while its
penumbral region is $\approx$230\,K cooler. This makes spot~C appear more
sun-like in terms of absolute penumbral temperature but its umbral contrast
with respect to the photosphere is only about a factor of three, compared to a
factor of $\sim$6--7 for our sunspot in Fig.~\ref{f:sunspot}. Moreover, spot~C
is elongated towards spot~B and looks more like a big solar active region than
a single spot. This suggests a violent interaction between the two features
which likely would distort or even fake a penumbral structure. Many smaller
unresolved spots between the two features could be an equally likely
explanation. Spot~A may also be just an unresolved compound of many smaller
spots, but it is isolated and relatively circular.

The spectral resolution of all previous data sets of mostly between $65,000$
and $77,000$ (approximately eight resolution elements across the stellar disk)
was likely insufficient to reveal a penumbral structure because the true
umbral contrast cannot be captured at this resolution; particularly because
EK\,Dra's low $v\sin i$ is already near the limit that permits the construction
of reliable temperature or brightness maps. For example,
\citet{2016MNRAS.459.4325M}
and
\citet{2015MNRAS.449....8W}
unsuccessfully attempted brightness mapping of HD\,35296 and $\tau$\,Boo,
both having a comparable $v\sin i$ of 15\,\kms\ but lacking EK\,Dra's spot
activity.

\subsection{Comparison to previous maps}

To date, five temperature maps and seven brightness maps of EK\,Dra have been
published at rather different epochs. Figure~\ref{F-mag} indicates these
epochs with respect to the long-term light curve. Because the photometry is
best fitted with a periodically modulated long-term variation, which we may
call a spot cycle (8.9 years, see Sect.~\ref{S44}), we can identify the cycle
phase $\varphi$ for all DI epochs. The very first temperature map was by
\citet{DI1}
from data taken in 1995 ($\varphi=2.19$, see Sect.~\ref{S44}). Three
subsequent maps were made by
\citet{2007A&A...472..887J}
from data taken in 2001--2002 (from $\varphi=2.91$ to $\varphi=3.03$), and 
another by
\citet{2009AIPC.1094..660J}
was made using data taken in 2007.56 ($\varphi=3.59$). The brightness maps by
\citet{2016A&A...593A..35R}
were obtained for epochs 2007.1 ($\varphi=3.53$) and 2012.1 ($\varphi=4.09$)
and by
\citet{2017MNRAS.465.2076W}
for five epochs between 2006 and 2012 (from $\varphi=3.51$ to $\varphi=4.09$).

Thus, maps now cover a full 20-year period. The surface activity at the times
of these maps, indicated by the full photometric amplitude of the rotational
modulation, was considerably different. This is illustrated in
Fig.~\ref{F-mag}. During the present mapping season ($\varphi=4.45$) the star 
showed a peak-to-peak photometric amplitude of up to 0\fm13 in $V$ while 
for example\ in 2012 ($\varphi=4.09$), during the mapping efforts of
\citet{2016A&A...593A..35R}
and
\citet{2017MNRAS.465.2076W},
it was roughly half of this. For 2012,
\citet{2017MNRAS.465.2076W}
reconstructed a small polar spot on EK\,Dra (and an even bigger one for Jan.
2008) whereas
\citet{2016A&A...593A..35R}
did not recover a polar spot at all. However, the high-latitude feature seen
in the maps ranging from the end of 2006 to early 2008 by
\citet{2017MNRAS.465.2076W}
is in agreement with the 2007 map presented by
\citet{2009AIPC.1094..660J}
which also has a large high-latitude spot reaching a latitude of over 80\degr.
As already discussed by
\citet{2017MNRAS.465.2076W},
the mid- to high-latitude spots seem to migrate polewards. Assuming that we
see the same high-latitude feature in all these maps, it has a lifetime
longer than one year, whereas the low-latitude features appear and disappear
within a month. Despite longer lifetimes, the high-latitude spots seem to show
evolution (spot coverage, intensity) on a timescale of months. Furthermore, as
our latest map shows, the polar region is not covered by spots all the time.

Because DR has only been measured from Stokes~$V$ but not from Stokes~$I$, we
briefly discuss here the current Zeeman-Doppler imaging (ZDI) literature. The
first ZDI maps of EK\,Dra were published by
\citet{2016A&A...593A..35R}.
They report mean field strengths of 66~G and 89~G for epochs 2007.1 and
2012.1, respectively. Shortly after that,
\citet{2017MNRAS.465.2076W}
published five more ZDI maps of EK\,Dra, although partially using the same
data as
\citet{2016A&A...593A..35R}.
While
\citet{2016A&A...593A..35R}
combined 2007 observations and produced one map,
\citet{2017MNRAS.465.2076W}
used those observations separately to create two maps. The 2012 map of both
teams is based on the same data set.
\citet{2017MNRAS.465.2076W}
reported mean field strengths ranging from 54~G to 92~G in agreement with the
previous results.

Common to all Doppler maps is the fact that there were always high-latitude
and low-latitude spots coexisting, with the high-latitude -- sometimes nearly
polar -- spot always being the more dominant feature. If true and systematic,
this would indicate the existence and dominance of a non-axisymmetric dynamo
component. However, no signs of a polarity reversal have been detected in the
mapping of EK\,Dra so far
\citep{2017MNRAS.465.2076W},
although one has to note that most of the ZDI maps so far are from the same
cycle, only the last one being obtained around the time we assume a new spot
cycle has just started (see, Sect.~\ref{S44}).

\subsection{Long-term photometric behaviour}\label{S44}

   \begin{figure}[!t]
   \centering
   \includegraphics[width=.5\textwidth]{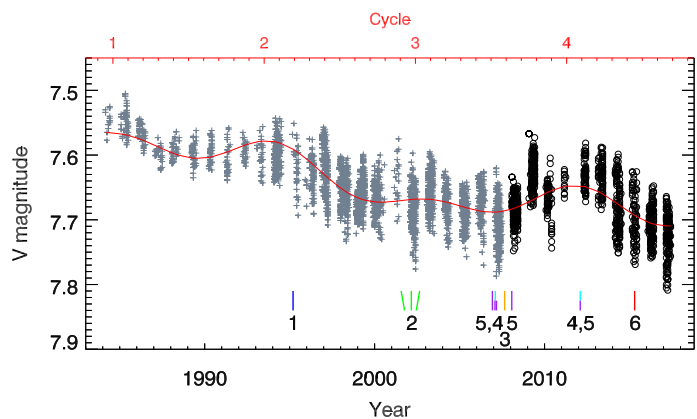}
   \caption{The long-term brightness record of EK\,Dra. The grey plus symbols
     denote previously published observations while the open black circles
     (from 2008 onward) represent new data points from this paper. The red
     line is the period fit to the data. The upper axis shows the cycle
     number/phase. The times for which there are published temperature or
     brightness maps are indicated with vertical ticks as follows:
     1 \citet{DI1},
     2 \citet{2007A&A...472..887J},
     3 \citet{2009AIPC.1094..660J},
     4 \citet{2016A&A...593A..35R},
     5 \citet{2017MNRAS.465.2076W},
     and 6 this paper.
              }
         \label{F-mag}
   \end{figure}

EK\,Dra has been photometrically monitored since the year 1958. For the first
$\sim$35~years, the monitoring was done using the Sonneberg Sky-Patrol
plates and was analysed by
\citet{2002A&A...391..659F}.
\citet{1997A&AS..125...11S}
analysed the photometry taken between 1994 and 1996 and reported an average
photometric period from those three seasons of 2.605 days, which was
interpreted to be the rotation period of EK\,Dra. They also noted the 
continuous decrease of the average $V$ light level. Later,
\citet{2003A&A...409.1017M}
analysed the long-term photometric record of the star until 2001 and concluded
that the modulation can be fitted by a sinusoid with a period of
$P_{\rm cyc} = 9.2$ years plus a longer-term trend ($P_{\rm cyc} \ge 30$ years
if cyclic). They also introduced for the first time a spot cycle for EK~Dra,
cycle~1 starting around 1984 and lasting until mid-1994. When all of the above
data were combined with results from more frequent monitoring
\citep{2005A&A...440..735J, 2009AIPC.1094..660J},
it was obvious that the star had been getting fainter for a time period of
$\sim$50~years, with some additional periodic variation of $\sim$10.5~years.
Adding eight more years of observations (Fig.~\ref{F-mag}) does not yet
confirm that the star has reached the minimum magnitude. For some time it
looked as though there was a minimum around 2006/2007. However, the latest
photometry confirmed that this was only a local and not a global minimum.

We have prewhitened the entire photometric record with the long-term trend of 
60+~years and refine the period to 8.9$\pm$0.2 years. The fit to the 
observations with the long-term cycle trend is shown in Fig.~\ref{F-mag}. The 
refined period is close to the period originally obtained by
\citet{2003A&A...409.1017M}
and is similar to the solar 11-year sunspot cycle, while the long-term trend
indicates the presence of a cycle that may be similar to the solar Gleissberg
cycle.

With the 8.9-year period, we can see that the photometric record in
Fig.~\ref{F-mag} already covers over four cycles. Cycle~1 started 1984.6, in
agreement with the definition given by
\citet{2003A&A...409.1017M},
corresponding to the time of photometric maximum (i.e. the smallest number
of spots on the stellar surface). The first modern photometric observations
(in late 1983) were taken at the end of cycle~0, just before cycle~1 began.
Based on the period of 8.9 years, cycle~0 must have started around 1975.7, 
and indeed, the analysis by
\citet{2002A&A...391..659F}
indicated that there was a photometric maximum around that time.


\section{Summary}

In this paper, we present the first temperature map based on
very-high-resolution spectra obtained with PEPSI at LBT. We show that the
relatively small line broadening of EK\,Dra, together with the only moderately
high spectral resolutions previously available, appear to be among the main
contributors to the lower-than-expected spot contrasts when comparing to the
Sun. A test inversion from degraded line profiles shows that the higher
spectral resolution of PEPSI reconstructs the surface with a 10\,\% higher 
temperature difference (on average) than before and with smaller surface areas
by $\sim$10-20\,\%.

The reconstructed stellar surface has four spots with temperature differences
between 990~ and 280~K below the photospheric temperature (5730$\pm$50~K).
Two of the spots are reconstructed with a typical solar morphology with an
umbra and a penumbra. For the one isolated and relatively round spot (spot A),
we determine an umbral temperature of 990\,K cooler than the unspotted
photosphere, whereas the penumbra is only 180\,K cooler. This leads to an
umbra to photosphere intensity ratio that is approximately only half of that
of our comparison sunspot. Spot~A of EK\,Dra is 27 times larger than this
comparison sunspot, and has  a diameter of $\sim$400\,Mm. There are no such
large spots on the Sun, but the EK\,Dra spot still appears much too warm in
its centre compared to its solar counterpart. We do not yet see evidence for a
conglomerate of little spots.

From the photometric record now covering almost 40 years, we have refined a
spot cycle period to $8.9\pm0.2$ years. Additionally, the photometry reveals
that the star still continues to fade.


\begin{acknowledgements}
  We thank the anonymous referee for careful reading of the manuscript and 
  very helpful comments.
  The LBT is an international collaboration among institutions in the United
  States, Italy, and Germany. LBT Corporation partners are the University of
  Arizona on behalf of the Arizona university system; Istituto Nazionale di
  Astrofisica, Italy; LBT Beteiligungsgesellschaft, Germany, representing
  the Max-Planck Society, the Leibniz-Institute for Astrophysics Potsdam
  (AIP), and Heidelberg University; the Ohio State University; and the
  Research Corporation, on behalf of the University of Notre Dame, University
  of Minnesota, and University of Virginia.
  It is a pleasure to thank the German Federal Ministry (BMBF) for the
  year-long support for the construction of PEPSI through their
  Verbundforschung grants 05AL2BA1/3 and 05A08BAC as well as the State of
  Brandenburg for the continuing support of AIP and PEPSI in particular
  (see https://pepsi.aip.de/).
  This work has made use of the VALD database, operated at Uppsala University,
  the Institute of Astronomy RAS in Moscow, and the University of Vienna.
  This research has made use of NASA's Astrophysics Data System and of CDS's
  Simbad database.
\end{acknowledgements}


   \bibliographystyle{aa} 
   \bibliography{EKPEPSI} 

\begin{thebibliography}{43}
\expandafter\ifx\csname natexlab\endcsname\relax\def\natexlab#1{#1}\fi

\bibitem[{{Allende Prieto} {et~al.}(2006){Allende Prieto}, {Beers}, {Wilhelm},
  {Newberg}, {Rockosi}, {Yanny}, \& {Lee}}]{2006ApJ...636..804A}
{Allende Prieto}, C., {Beers}, T.~C., {Wilhelm}, R., {et~al.} 2006, \apj, 636,
  804

\bibitem[{{Ayres} \& {France}(2010)}]{2010ApJ...723L..38A}
{Ayres}, T. \& {France}, K. 2010, \apjl, 723, L38

\bibitem[{{Ayres}(2015)}]{2015AAS...22513823A}
{Ayres}, T.~R. 2015, in American Astronomical Society Meeting Abstracts, Vol.
  225, 138.23

\bibitem[{{Balthasar}(2006)}]{sunspot}
{Balthasar}, H. 2006, \aap, 449, 1169

\bibitem[{{Berdyugina} {et~al.}(2003){Berdyugina}, {Telting}, \&
  {Korhonen}}]{2003A&A...406..273B}
{Berdyugina}, S.~V., {Telting}, J.~H., \& {Korhonen}, H. 2003, \aap, 406, 273

\bibitem[{{Bray} \& {Loughhead}(1964)}]{1964suns.book.....B}
{Bray}, R.~J. \& {Loughhead}, R.~E. 1964, {Sunspots} (Chapman and Hall, London)

\bibitem[{{Carroll} {et~al.}(2007){Carroll}, {Kopf}, {Ilyin}, \&
  {Strassmeier}}]{2007AN....328.1043C}
{Carroll}, T.~A., {Kopf}, M., {Ilyin}, I., \& {Strassmeier}, K.~G. 2007,
  Astronomische Nachrichten, 328, 1043

\bibitem[{{Carroll} {et~al.}(2009){Carroll}, {Kopf}, {Strassmeier}, \&
  {Ilyin}}]{2009IAUS..259..633C}
{Carroll}, T.~A., {Kopf}, M., {Strassmeier}, K.~G., \& {Ilyin}, I. 2009, in IAU
  Symposium, ed. K.~G. {Strassmeier}, A.~G. {Kosovichev}, \& J.~E. {Beckman},
  Vol. 259, 633

\bibitem[{{Carroll} {et~al.}(2012){Carroll}, {Strassmeier}, {Rice}, \&
  {K{\"u}nstler}}]{2012A&A...548A..95C}
{Carroll}, T.~A., {Strassmeier}, K.~G., {Rice}, J.~B., \& {K{\"u}nstler}, A.
  2012, \aap, 548, A95

\bibitem[{{Castelli} \& {Kurucz}(2004)}]{2004astro.ph..5087C}
{Castelli}, F. \& {Kurucz}, R.~L. 2004, ArXiv Astrophysics e-prints
  [\eprint{astro-ph/0405087}]

\bibitem[{{Collier Cameron}(1992)}]{1992LNP...397...33C}
{Collier Cameron}, A. 1992, in Lecture Notes in Physics, Berlin Springer
  Verlag, Vol. 397, Surface Inhomogeneities on Late-Type Stars, ed. P.~B.
  {Byrne} \& D.~J. {Mullan}, 33

\bibitem[{{Dorren} \& {Guinan}(1994)}]{1994ApJ...428..805D}
{Dorren}, J.~D. \& {Guinan}, E.~F. 1994, \apj, 428, 805

\bibitem[{{Dravins}(2008)}]{2008A&A...492..199D}
{Dravins}, D. 2008, \aap, 492, 199

\bibitem[{{Flores Soriano} \& {Strassmeier}(2017)}]{2017A&A...597A.101F}
{Flores Soriano}, M. \& {Strassmeier}, K.~G. 2017, \aap, 597, A101

\bibitem[{{Fr{\"o}hlich} {et~al.}(2002){Fr{\"o}hlich}, {Tsch{\"a}pe},
  {R{\"u}diger}, \& {Strassmeier}}]{2002A&A...391..659F}
{Fr{\"o}hlich}, H.-E., {Tsch{\"a}pe}, R., {R{\"u}diger}, G., \& {Strassmeier},
  K.~G. 2002, \aap, 391, 659

\bibitem[{{Gaia Collaboration} {et~al.}(2016){Gaia Collaboration}, {Brown},
  {Vallenari}, {Prusti}, {de Bruijne}, {Mignard}, {Drimmel}, {Babusiaux},
  {Bailer-Jones}, {Bastian}, \& et~al.}]{2016A&A...595A...2G}
{Gaia Collaboration}, {Brown}, A.~G.~A., {Vallenari}, A., {et~al.} 2016, \aap,
  595, A2

\bibitem[{{Granzer} {et~al.}(2001){Granzer}, {Reegen}, \&
  {Strassmeier}}]{2001AN....322..325G}
{Granzer}, T., {Reegen}, P., \& {Strassmeier}, K.~G. 2001, Astronomische
  Nachrichten, 322, 325

\bibitem[{{Gustafsson}(2007)}]{2007ASPC..378...60G}
{Gustafsson}, B. 2007, in Astronomical Society of the Pacific Conference
  Series, Vol. 378, Why Galaxies Care About AGB Stars: Their Importance as
  Actors and Probes, ed. F.~{Kerschbaum}, C.~{Charbonnel}, \& R.~F. {Wing}, 60

\bibitem[{{Ilyin}(2000)}]{4A}
{Ilyin}, I.~V. 2000, PhD thesis, Astronomy Division Department of Physical
  Sciences P.O.Box 3000 FIN-90014 University of Oulu Finland

\bibitem[{{J{\"a}rvinen} {et~al.}(2007){J{\"a}rvinen}, {Berdyugina},
  {Korhonen}, {Ilyin}, \& {Tuominen}}]{2007A&A...472..887J}
{J{\"a}rvinen}, S.~P., {Berdyugina}, S.~V., {Korhonen}, H., {Ilyin}, I., \&
  {Tuominen}, I. 2007, \aap, 472, 887

\bibitem[{{J{\"a}rvinen} {et~al.}(2005){J{\"a}rvinen}, {Berdyugina}, \&
  {Strassmeier}}]{2005A&A...440..735J}
{J{\"a}rvinen}, S.~P., {Berdyugina}, S.~V., \& {Strassmeier}, K.~G. 2005, \aap,
  440, 735

\bibitem[{{J{\"a}rvinen} {et~al.}(2009){J{\"a}rvinen}, {Korhonen},
  {Berdyugina}, \& {Ilyin}}]{2009AIPC.1094..660J}
{J{\"a}rvinen}, S.~P., {Korhonen}, H., {Berdyugina}, S.~V., \& {Ilyin}, I.
  2009, in American Institute of Physics Conference Series, Vol. 1094, 15th
  Cambridge Workshop on Cool Stars, Stellar Systems, and the Sun, ed.
  E.~{Stempels}, 660

\bibitem[{{Jofr{\'e}} {et~al.}(2014){Jofr{\'e}}, {Heiter}, {Soubiran},
  {Blanco-Cuaresma}, {Worley}, {Pancino}, {Cantat-Gaudin}, {Magrini},
  {Bergemann}, {Gonz{\'a}lez Hern{\'a}ndez}, {Hill}, {Lardo}, {de Laverny},
  {Lind}, {Masseron}, {Montes}, {Mucciarelli}, {Nordlander}, {Recio Blanco},
  {Sobeck}, {Sordo}, {Sousa}, {Tabernero}, {Vallenari}, \& {Van
  Eck}}]{2014A&A...564A.133J}
{Jofr{\'e}}, P., {Heiter}, U., {Soubiran}, C., {et~al.} 2014, \aap, 564, A133

\bibitem[{{K{\"u}nstler} {et~al.}(2015){K{\"u}nstler}, {Carroll}, \&
  {Strassmeier}}]{XXTri}
{K{\"u}nstler}, A., {Carroll}, T.~A., \& {Strassmeier}, K.~G. 2015, \aap, 578,
  A101

\bibitem[{{Kupka} {et~al.}(2011){Kupka}, {Dubernet}, \& {VAMDC
  Collaboration}}]{2011BaltA..20..503K}
{Kupka}, F., {Dubernet}, M.-L., \& {VAMDC Collaboration}. 2011, Baltic
  Astronomy, 20, 503

\bibitem[{{K{\"u}rster}(1993)}]{1993A&A...274..851K}
{K{\"u}rster}, M. 1993, \aap, 274, 851

\bibitem[{{Mengel} {et~al.}(2016){Mengel}, {Fares}, {Marsden}, {Carter},
  {Jeffers}, {Petit}, {Donati}, {Folsom}, \& {BCool
  Collaboration}}]{2016MNRAS.459.4325M}
{Mengel}, M.~W., {Fares}, R., {Marsden}, S.~C., {et~al.} 2016, \mnras, 459,
  4325

\bibitem[{{Messina} \& {Guinan}(2003)}]{2003A&A...409.1017M}
{Messina}, S. \& {Guinan}, E.~F. 2003, \aap, 409, 1017

\bibitem[{{Piskunov} \& {Rice}(1993)}]{1993PASP..105.1415P}
{Piskunov}, N.~E. \& {Rice}, J.~B. 1993, \pasp, 105, 1415

\bibitem[{{Piskunov} \& {Wehlau}(1990)}]{1990A&A...233..497P}
{Piskunov}, N.~E. \& {Wehlau}, W.~H. 1990, \aap, 233, 497

\bibitem[{{Rice} \& {Strassmeier}(2000)}]{2000A&AS..147..151R}
{Rice}, J.~B. \& {Strassmeier}, K.~G. 2000, \aaps, 147, 151

\bibitem[{{Ros{\'e}n} {et~al.}(2016){Ros{\'e}n}, {Kochukhov}, {Hackman}, \&
  {Lehtinen}}]{2016A&A...593A..35R}
{Ros{\'e}n}, L., {Kochukhov}, O., {Hackman}, T., \& {Lehtinen}, J. 2016, \aap,
  593, A35

\bibitem[{{Ryabchikova} {et~al.}(2015){Ryabchikova}, {Piskunov}, {Kurucz},
  {Stempels}, {Heiter}, {Pakhomov}, \& {Barklem}}]{2015PhyS...90e4005R}
{Ryabchikova}, T., {Piskunov}, N., {Kurucz}, R.~L., {et~al.} 2015, Phys.\ Scr.,
  90, 054005

\bibitem[{{Strassmeier}(2009)}]{2009A&ARv..17..251S}
{Strassmeier}, K.~G. 2009, \aapr, 17, 251

\bibitem[{{Strassmeier} {et~al.}(1997{\natexlab{a}}){Strassmeier}, {Bartus},
  {Cutispoto}, \& {Rodono}}]{1997A&AS..125...11S}
{Strassmeier}, K.~G., {Bartus}, J., {Cutispoto}, G., \& {Rodono}, M.
  1997{\natexlab{a}}, \aaps, 125, 11

\bibitem[{{Strassmeier} {et~al.}(1997{\natexlab{b}}){Strassmeier}, {Boyd},
  {Epand}, \& {Granzer}}]{1997PASP..109..697S}
{Strassmeier}, K.~G., {Boyd}, L.~J., {Epand}, D.~H., \& {Granzer}, T.
  1997{\natexlab{b}}, \pasp, 109, 697

\bibitem[{{Strassmeier} {et~al.}(2015){Strassmeier}, {Ilyin}, {J{\"a}rvinen},
  {Weber}, {Woche}, {Barnes}, {Bauer}, {Beckert}, {Bittner}, {Bredthauer},
  {Carroll}, {Denker}, {Dionies}, {DiVarano}, {D{\"o}scher}, {Fechner},
  {Feuerstein}, {Granzer}, {Hahn}, {Harnisch}, {Hofmann}, {Lesser}, {Paschke},
  {Pankratow}, {Plank}, {Pl{\"u}schke}, {Popow}, \&
  {Sablowski}}]{2015AN....336..324S}
{Strassmeier}, K.~G., {Ilyin}, I., {J{\"a}rvinen}, A., {et~al.} 2015,
  Astronomische Nachrichten, 336, 324

\bibitem[{{Strassmeier} {et~al.}(2018{\natexlab{a}}){Strassmeier}, {Ilyin}, \&
  {Steffen}}]{2017Strassmeier}
{Strassmeier}, K.~G., {Ilyin}, I., \& {Steffen}, M. 2018{\natexlab{a}}, \aap,
  612, A44

\bibitem[{{Strassmeier} {et~al.}(2018{\natexlab{b}}){Strassmeier}, {Ilyin}, \&
  {Weber}}]{2017PEPSI}
{Strassmeier}, K.~G., {Ilyin}, I., \& {Weber}, M. 2018{\natexlab{b}}, \aap,
  612, A45

\bibitem[{{Strassmeier} \& {Rice}(1998)}]{DI1}
{Strassmeier}, K.~G. \& {Rice}, J.~B. 1998, \aap, 330, 685

\bibitem[{{Vogt} \& {Penrod}(1983)}]{1983PASP...95..565V}
{Vogt}, S.~S. \& {Penrod}, G.~D. 1983, \pasp, 95, 565

\bibitem[{{Waite} {et~al.}(2015){Waite}, {Marsden}, {Carter}, {Petit},
  {Donati}, {Jeffers}, \& {Boro Saikia}}]{2015MNRAS.449....8W}
{Waite}, I.~A., {Marsden}, S.~C., {Carter}, B.~D., {et~al.} 2015, \mnras, 449,
  8

\bibitem[{{Waite} {et~al.}(2017){Waite}, {Marsden}, {Carter}, {Petit},
  {Jeffers}, {Morin}, {Vidotto}, {Donati}, \& {BCool
  Collaboration}}]{2017MNRAS.465.2076W}
{Waite}, I.~A., {Marsden}, S.~C., {Carter}, B.~D., {et~al.} 2017, \mnras, 465,
  2076

\end{thebibliography}
\end{document}